\pdfoutput=1

\documentclass{pic2012}
\def\runauthor{Francesco Santanastasio}
\def\shorttitle{Exotic Phenomena Searches at Hadron Colliders}
\usepackage{xspace}

\def\invfb {fb$^{-1}$\xspace}
\def\etmiss {\ensuremath{E_{\mathrm{T}}\hspace{-1.1em}/\kern0.5em}\xspace}
\def\pt{\ensuremath{p_{\rm T}}\xspace}
\def\koverm{$k/\bar{M}_{\rm{Planck}}$\xspace}
\def\eejj{$\rm{eejj}$\xspace}
\def\mumujj{$\mu\mu \rm{jj}$\xspace}
\def\enujj{$\rm{e}\nu \rm{jj}$\xspace}
\def\munujj{$\mu\nu \rm{jj}$\xspace}
\def\nunubb{$\nu\nu \rm{bb}$\xspace}
\def\tautaubb{$\tau\tau \rm{bb}$\xspace}

\begin{document}

\title{Exotic Phenomena Searches at Hadron Colliders}

\author{Francesco Santanastasio \\ (on behalf of the ATLAS and CMS Collaborations)}

\address{European Organization for Nuclear Research \\ 
CERN CH-1211, Geneve 23, Switzerland \\
E-mail: francesco.santanastasio@cern.ch }

\maketitle

\abstracts{This review presents a selection of the final results of searches for 
various exotic physics phenomena in proton-proton collisions at $\sqrt{s}=7$ and 8~TeV delivered 
by the LHC and collected with the ATLAS and CMS detectors in 2011 (5~\invfb) 
and in the first part of 2012 (4~\invfb). 
%The data corresponds to an integrated luminosity of 5~\invfb. 
Searches for large extra dimensions, gravitons, microscopic black holes, long-lived particles, 
dark matter, and leptoquarks are presented in this report. 
%A search for microscopic black holes 
%performed with 3.7~\invfb of proton-proton collisions at $\sqrt{s}=8$~TeV collected in 2012 by the 
%CMS detector is also shown. 
No sign of new physics beyond the Standard Model has been observed so far. 
In the majority of the cases these searches set the most stringent 
limits to date on the aforementioned new physics phenomena.}

\section{Introduction}
A selection of searches for various exotic physics phenomena beyond 
the Standard Model (SM) in proton-proton collisions at $\sqrt{s}=7$~TeV delivered by the LHC and collected with the 
ATLAS~\cite{Aad:2008zzm} and CMS~\cite{Chatrchyan:2008aa} detectors in 2011 is presented in this paper. 
The data correspond to an integrated luminosity of 5~\invfb. 
Searches for large extra dimensions, gravitons, long-lived particles, 
dark matter, and leptoquarks are presented in this report. A preliminary result 
of a search for microscopic black holes performed with 3.7~\invfb of proton-proton collisions 
at $\sqrt{s}=8$~TeV collected in 2012 by the CMS detector is also shown.
Searches for Supersymmetry (SUSY) and other exotic physics phenomena (such as 
searches for new heavy fermions and bosons) are not discussed in this paper. 
These results can be found in other proceedings of this conference.
The complete set of public results from ATLAS and CMS experiments can be 
found in references~\cite{ATLASpage,CMSpage}.

\section{Large Extra Dimensions}
Compact large extra dimensions (ED) are an intriguing proposed solution to the hierarchy 
problem of the SM, which refers to the puzzling fact that 
the fundamental scale of gravity $M_{\rm{Planck}} \sim 10^{19}$~GeV is so much higher 
than the electroweak symmetry breaking scale~$\sim 10^3$~GeV. In the ADD 
model~\footnote{The original proposal to use large extra dimensions to solve the hierarchy problem was presented 
by Arkani-Hamed, Dimopoulos, and Dvali (ADD).}, the SM is constrained to the common 
3+1 space-time dimensions, while gravity is free to propagate through the entire 
multidimensional space. The gravitational flux in 3+1 dimensions is effectively diluted by 
virtue of the multidimensional Gauss's Law. In this framework, the fundamental Planck scale 
can be lowered to the electroweak scale, thus making production of gravitons possible at the LHC.
Experimental signatures indicative for the existence of EDs are discussed below.
%Some of the experimental signatures of the existence of such extra dimensions are discussed below.

\subsection{Microscopic Black Holes (8 TeV)}
\begin{figure}[htbp]
%\vspace*{7.0cm}
\begin{center}
%\special{psfile=pic2012_template_fig.ps voffset=-60 vscale=40
%hscale= 40 hoffset=10 angle=0}
    \begin{tabular}{cc}
    \psfig{figure=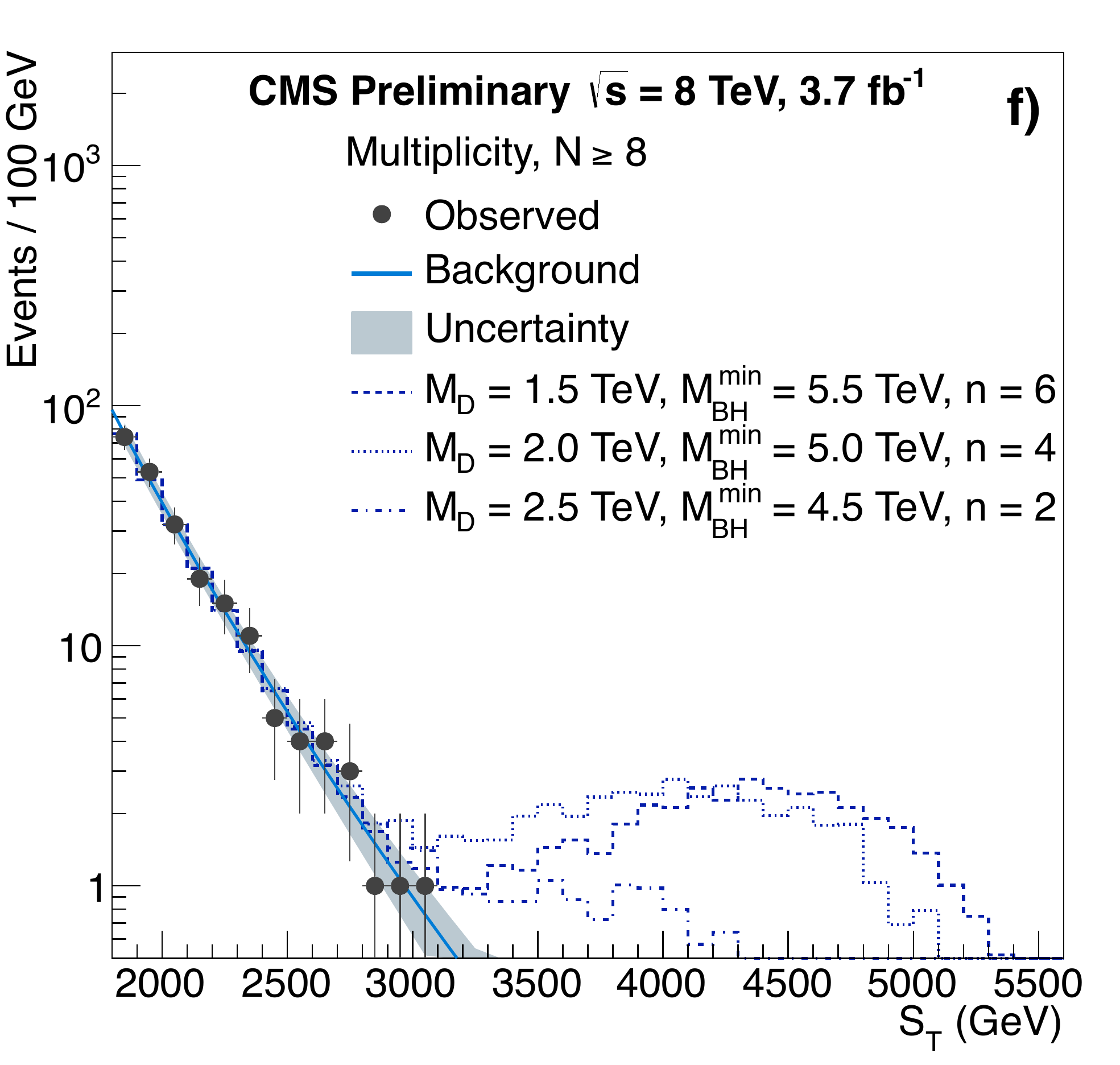,height=1.8in} &
      \psfig{figure=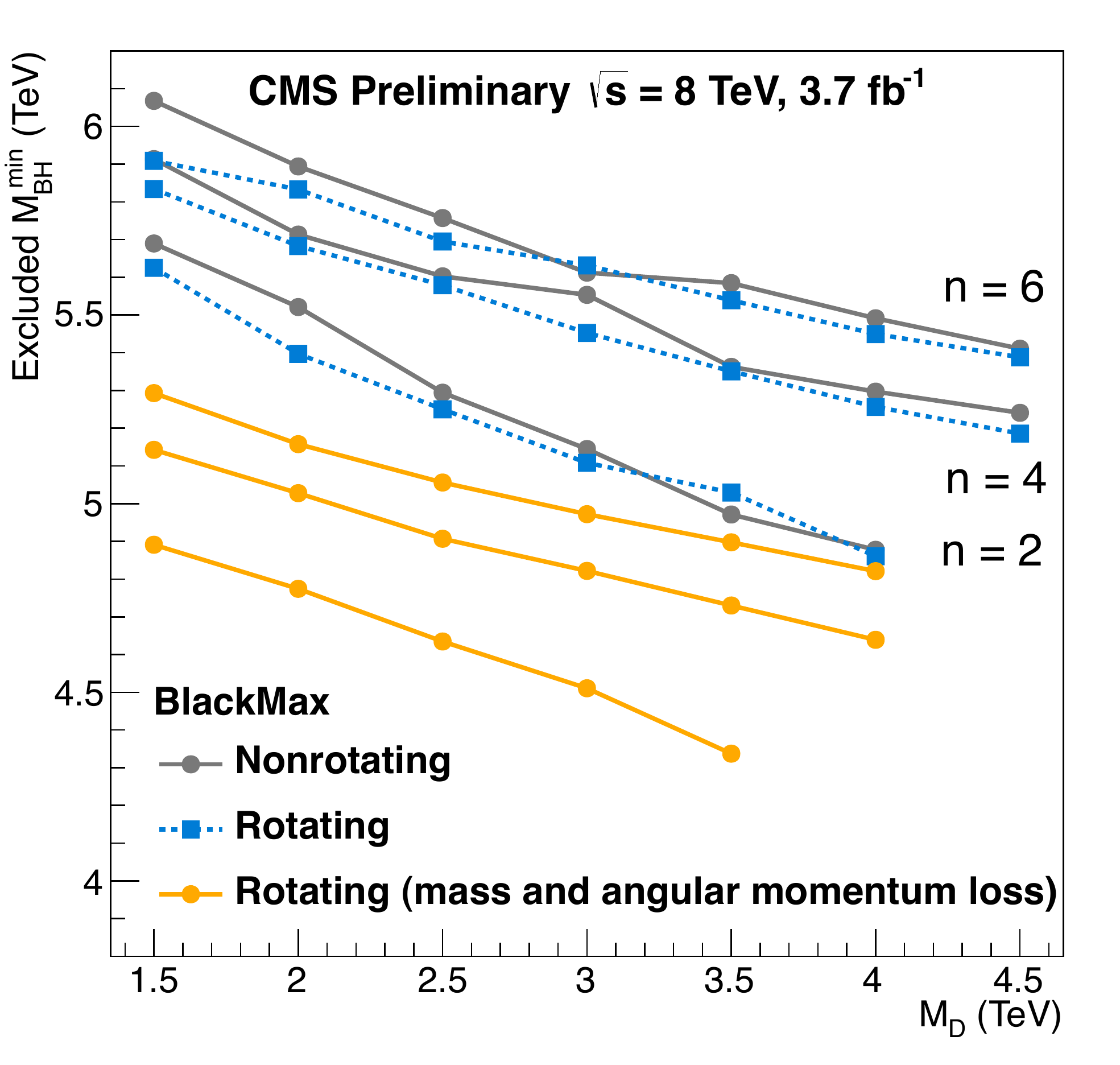,height=1.8in} \\
%\centerline{\epsfxsize=2.9in\epsfbox{BlakHoleST_CMS.pdf}} &
%\centerline{\epsfxsize=2.9in\epsfbox{BlakHoleST_CMS.pdf}} \\
    \end{tabular}
\caption[*]{ (Left) Total transverse energy (including the missing transverse energy in the sum) 
for events with at least 8 objects (jets, leptons, and photons) for data, background prediction, and black hole signals for three different parameter sets. (Right) Minimum black hole mass excluded at 95\% CL as function of the reduced Planck scale for various {\sc BlackMax} black hole models without the stable remnant and number of extra dimensions of two, four, and six~\cite{CMS-PAS-EXO-12-009-BlackHoles}.}
\label{fig:BlackHole}    
\end{center}
\end{figure}

One of the exciting predictions of theoretical models with extra dimensions 
and low-scale quantum gravity is the possibility of copious production of 
microscopic black holes in particle collisions at the LHC.
Events with large total transverse energy are analyzed for the presence
of multiple high-energy jets, leptons, and photons, typical of a signal expected 
from a microscopic black hole. 
The analysis is performed using the first 3.7~\invfb of data collected by the 
CMS experiment in 2012 at $\sqrt{s}=8$~TeV~\cite{CMS-PAS-EXO-12-009-BlackHoles}. 
Figure~\ref{fig:BlackHole} (left) shows the distribution of the total transverse energy for data, 
background prediction and various signal samples. Good agreement with the Standard Model backgrounds, 
dominated by QCD multijet production, is observed for various final-state multiplicities 
and model-independent limits on new physics in these final states are set. 
Using a simple semi-classical approximation, new model-specific indicative limits 
on the minimum black hole mass are derived as well in the range 4.1 -- 6.1~TeV, depending 
on the specific model considered. An example of limits using the {\sc BlackMax} generator is reported 
in Fig.~\ref{fig:BlackHole} (right).
The analysis has a substantially increased sensitivity compared to previous searches 
due to higher collision energy.

%\subsection{Mono-Jet and Mono-Photon Final States (including Dark Matter Limits)}
\subsection{Real Graviton Production and Dark Matter Limits}
\begin{figure}[htbp]
%\vspace*{7.0cm}
\begin{center}
%\special{psfile=pic2012_template_fig.ps voffset=-60 vscale=40
%hscale= 40 hoffset=10 angle=0}
    \begin{tabular}{cc}
    \psfig{figure=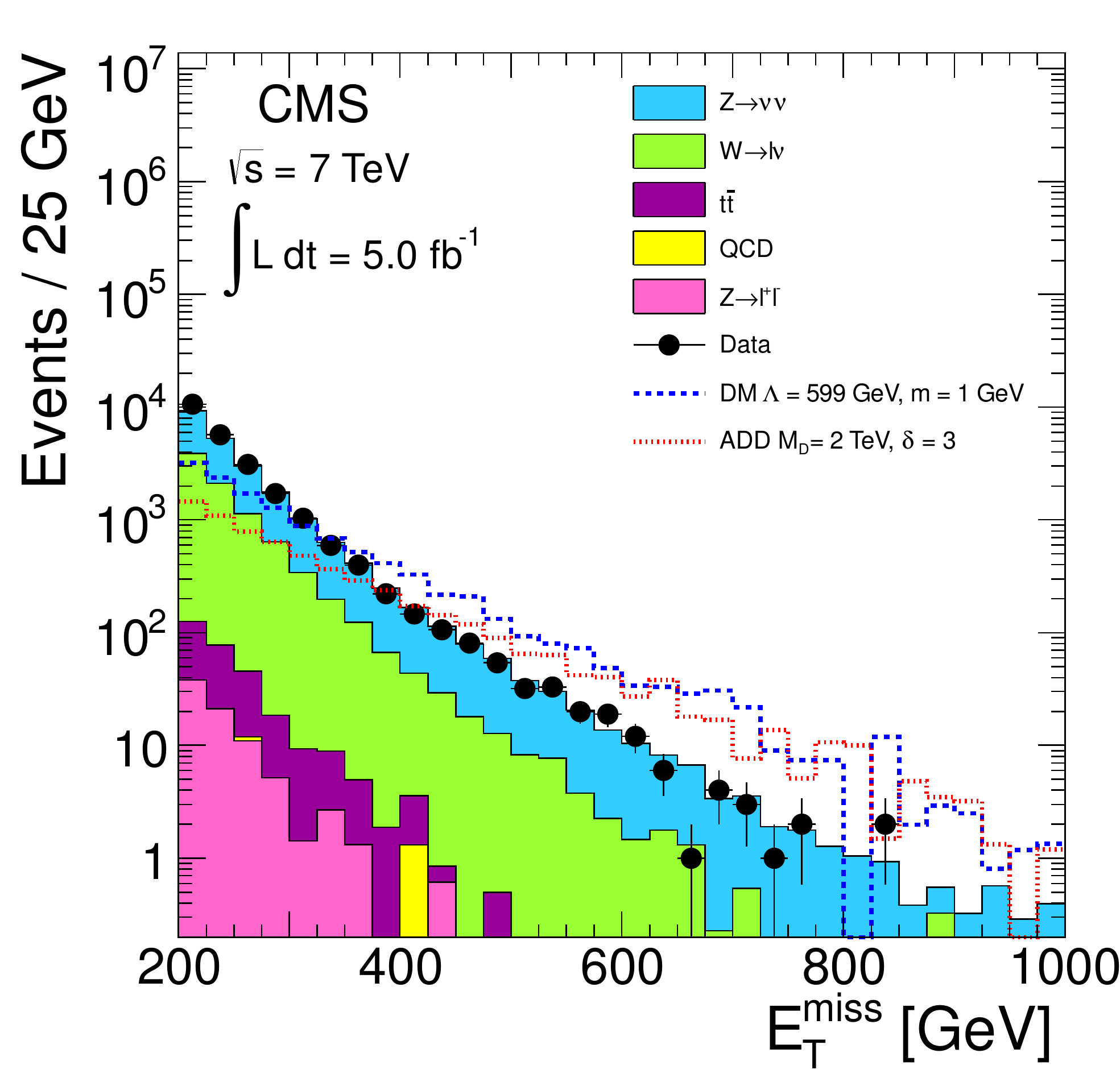,height=1.55in} &
    \psfig{figure=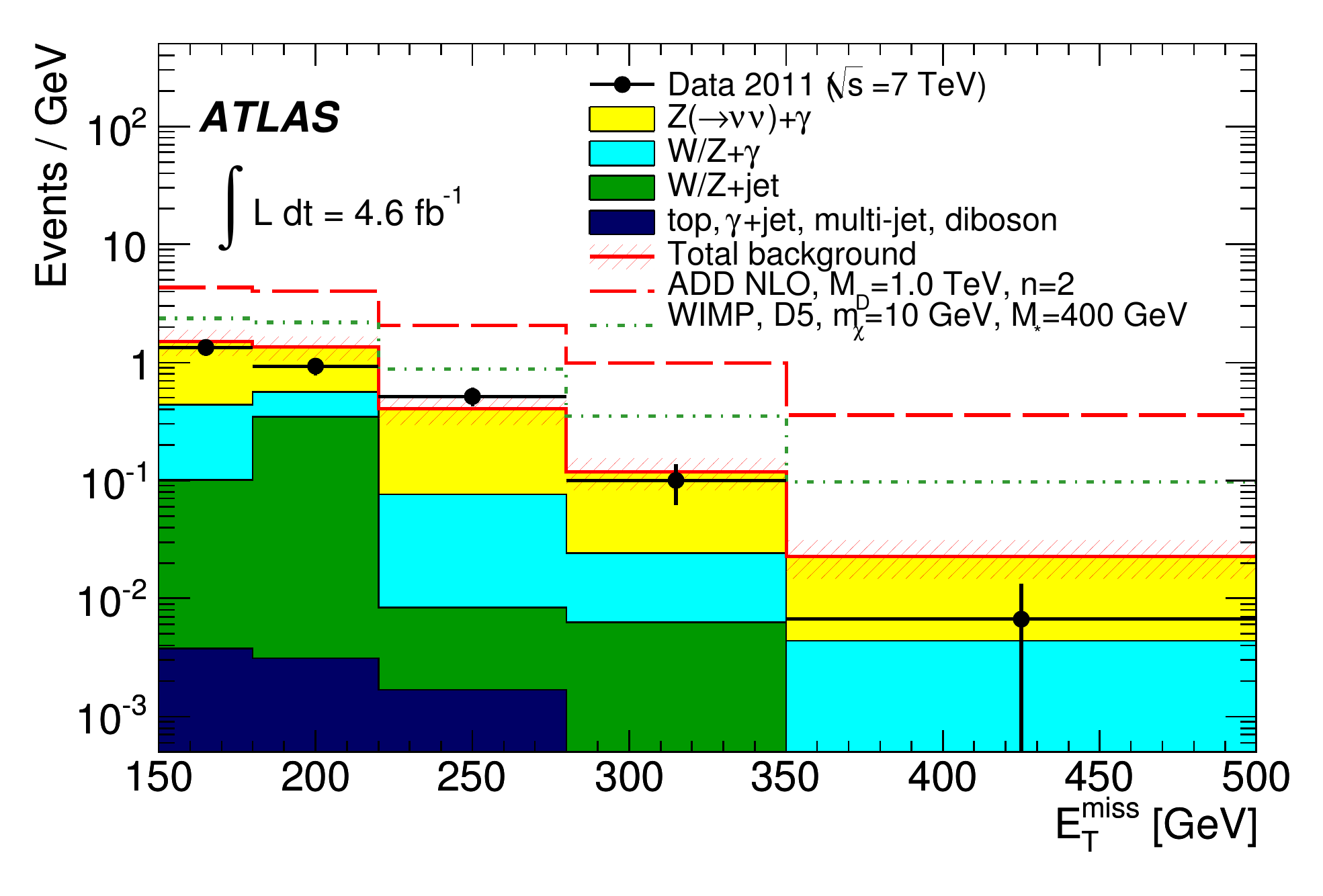,height=1.55in} \\
    \psfig{figure=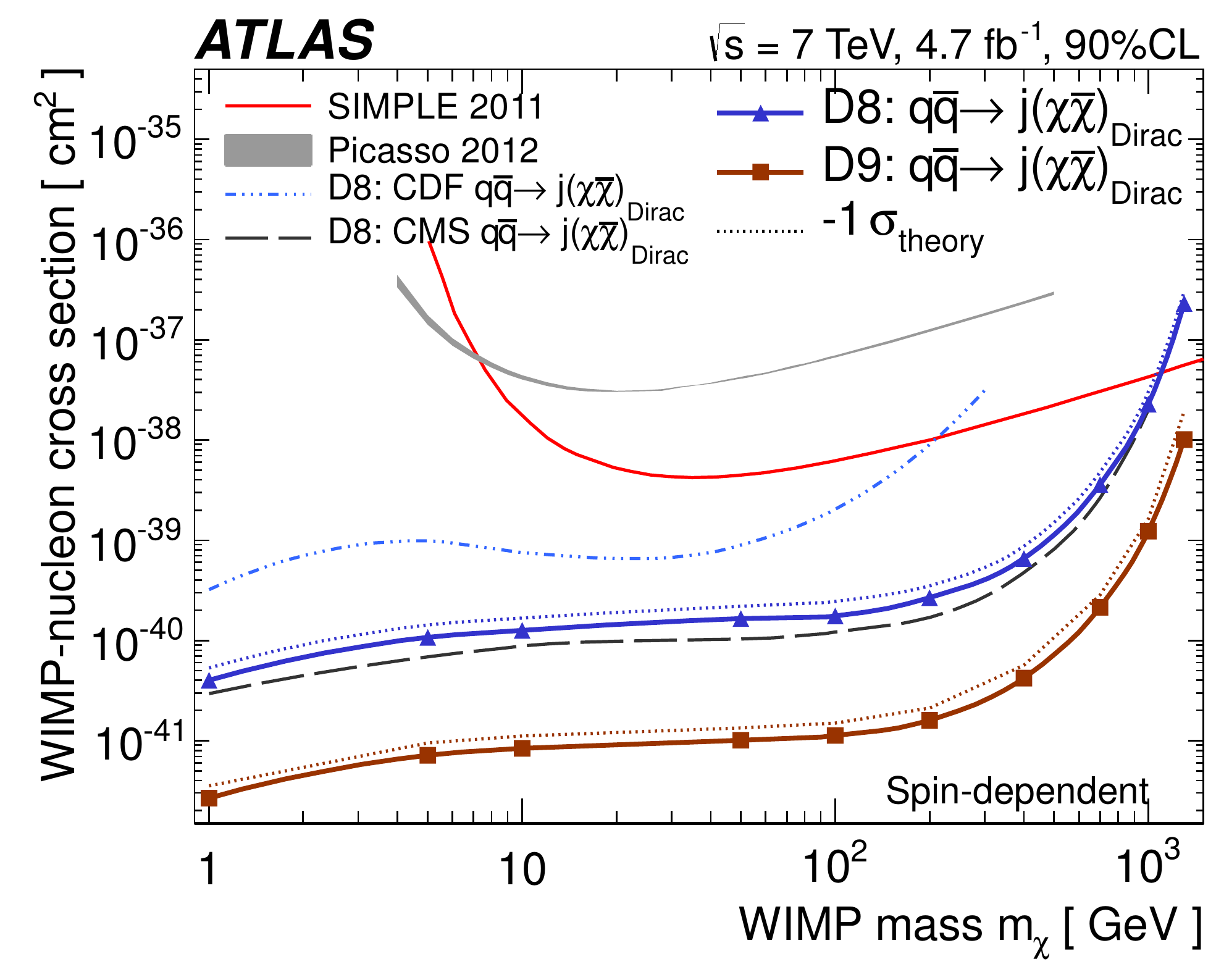,height=1.6in} &
    \psfig{figure=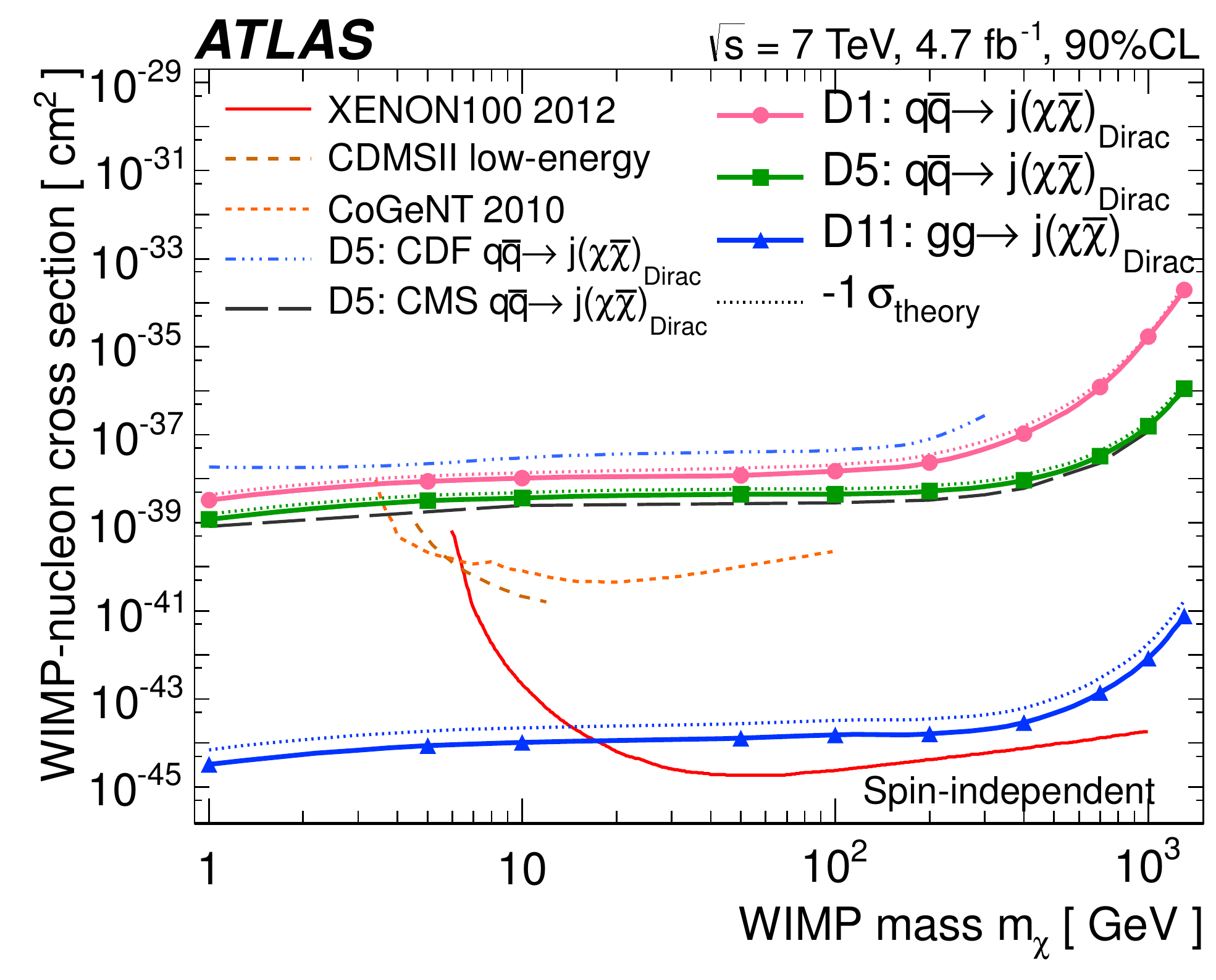,height=1.6in} \\
%\centerline{\epsfxsize=2.9in\epsfbox{BlakHoleST_CMS.pdf}} &
%\centerline{\epsfxsize=2.9in\epsfbox{BlakHoleST_CMS.pdf}} \\
    \end{tabular}
\caption[*]{(Top) Distributions of \etmiss for data and SM background predictions
after the CMS mono-jet (top-left)~\cite{Chatrchyan:2012me} 
and ATLAS mono-photon (top-right)~\cite{Aad:2012fw} event selection. 
Examples of ADD and dark matter (indicated as DM or WIMP) signals are also shown.
(Bottom) Limits on spin-dependent (bottom-left) and spin-independent (bottom-right) WIMP-nucleon
cross section as a function of WIMP mass from ATLAS and CMS mono-jet searches compared to direct and indirect detection experiments~\cite{ATLAS:2012ky}. 
Limits derived from mono-photon searches are slightly weaker.}
\label{fig:MonoJetAndMonoPhoton}    
\end{center}
\end{figure}

Searches for production of a real graviton produced in association with either an 
energetic hadronic jet or a photon have been performed by both ATLAS~\cite{ATLAS:2012ky,Aad:2012fw} and CMS~\cite{Chatrchyan:2012me,Chatrchyan:2012tea} experiments.
Since gravitons are free to propagate in the extra dimensions, they escape 
the detector and can only be inferred from the amount of missing transverse energy (\etmiss). 
The offline event selection requires large \etmiss, one high \pt jet or photon, a veto on the 
presence of well-identified leptons and isolated tracks, and additional 
requirements to suppress the cosmics, beam halo, and instrumental backgrounds 
that can fake the mono-jet+\etmiss and mono-photon+\etmiss signatures.
Figure~\ref{fig:MonoJetAndMonoPhoton} (top) shows the \etmiss distributions 
in mono-jet and mono-photon events from the CMS and ATLAS analyses, respectively.
The number of observed events in data is in good agreement with the SM prediction, 
and significant improvements are made to the existing limits on the fundamental parameters of 
the extra dimension model describing real-graviton emission.

These analyses are also sensitive to pair production of weakly interacting dark matter 
particles (WIMP, denoted as $\chi$) where the $\chi$-$\chi$ system recoils 
against an energetic hadronic jet or photon from initial state radiation. 
Using an effective field theory,  the experimental collider results can be used to derive limits on the 
WIMP-nucleon cross section as a function of the mass of the dark matter candidate
as shown in Fig.~\ref{fig:MonoJetAndMonoPhoton} (bottom) for the mono-jet analyses.
Under the assumptions of this model, the ATLAS and CMS limits are more stringent than the ones from direct and indirect detection experiments for the spin-dependent WIMP-nucleon scattering over the entire WIMP mass ($m_{\chi}$) range. For the spin-independent scattering the collider limits are the most stringent ones for $m_{\chi}<10$~GeV.

\section{Warped Extra Dimensions}

Another possible solution to the hierarchy problem of the SM 
is based on the original Randall-Sundrum framework (RS1) 
with a warped extra dimension~\cite{Randall:1999ee}. The most 
distinctive novel feature of this scenario is the existence of spin-2 Kaluza-Klein 
gravitons whose masses and couplings to the SM are set by the TeV scale. 
These gravitons would appear in experiments as widely-sephighlyarated-in-mass resonances, 
in contrast to the very light, closely-spaced-in-mass gravitons predicted in large extra 
dimension models. In the RS1 scenario, decays of gravitons to pairs of photons, 
leptons, or light jets provides the most striking experimental signatures for discovery 
of this new particle. The two model parameters are the graviton mass and the coupling \koverm, 
where $k$ is the curvature scale of the warped extra dimension and $\bar{M}_{\rm{Planck}}$
is the reduced Planck mass. Upper limits on \koverm were derived as a function of the graviton mass 
more than 10 years ago by using electroweak precision measurements (oblique parameters S, T)~\cite{Davoudiasl:2000wi}. These studies demand that \koverm be less than $\approx 0.1$.

A well-motivated extension of the original RS1 model (bulk graviton)~\cite{Agashe:2007zd}
addresses the flavor structure of the SM through localization of fermions in the warped bulk
of the theory. This picture offers a unified geometric explanation of both the hierarchy 
and the flavor puzzles in the SM. In this case, graviton production and decay via light fermions
is highly suppressed and the decay into photons is negligible. On the other hand,  
production of bulk gravitons from gluon fusion and their decay into longitudinally polarized 
gauge bosons W/Z can be significant. Depending on the model parameters, production of gravitons via vector boson fusion (VBF) can be sizeable as well. Recent theoretical studies done in the contest of the bulk graviton model show that values of \koverm as large as $\sim 3$ are still within the validity of 
the model~\cite{Agashe:2007zd}. It should be noted that for values of \koverm greater than 
2 the bulk graviton width becomes larger than $\sim 20$\% of its mass, thus introducing 
experimental issues in the detection of such resonances which are not discussed in this paper.

\subsection{Examples of RS1 Graviton Signatures}

Both ATLAS and CMS collaborations have searched for narrow resonances  
in the invariant mass spectrum of dielectron/dimuon~\cite{Aad:2012hf,Chatrchyan:2012it}  
and diphoton~\cite{Aad:2012cy,Chatrchyan:2011fq} final states. The spectra are consistent with 
the SM expectations in both the bulk and the tails of the aforementioned 
distributions. Among those searches, the most stringent lower limits on the mass of RS1 gravitons 
are 0.92 (2.16)~TeV for $\mbox{\koverm}=0.01$ (0.1) in the dielectron$+$dimuon channel,
and 1 (2.06)~TeV for $\mbox{\koverm}=0.01$ (0.1) in the diphoton channel.

\subsection{Examples of Bulk Graviton Signatures}

Various searches for heavy, exotic resonances (X) decaying into 
pairs of vector bosons (W/Z) have been performed by both ATLAS and CMS experiments.
The higher the mass of the resonance the larger the boost of its decay products, the W/Z bosons;
consequently, their decay products tend to be close in 
%Higher is the mass of the resonance and more the decay products of the W/Z bosons 
%will be boosted in the laboratory frame. 
%Leptons tend to be close in 
$\Delta R=\sqrt{\Delta\phi^2+\Delta\eta^2}$ thus requiring special 
algorithms to define lepton isolation and identification criteria.
Jets from hadronic W/Z decays might also merge into a single wide jet (W/Z-jet) thus requiring the 
use of jet substructure techniques to distinguish W/Z-jets from regular QCD-jets.
The transition between boosted and un-boosted event decay topologies
(the latter represents the case where all the W/Z decay products are typically reconstructed as well separated objects in the ATLAS and CMS detectors) happens around a mass of 1 TeV for resonances decaying to a pair of W/Z bosons. 
The analyses of ZZ and WW final states performed by ATLAS and CMS using 2011 data at $\sqrt{s}=7$~TeV
in the contest of exotic phenomena searches (i.e. excluding searches for SM Higgs boson) are: \\ \\ 
%\begin{itemize} 
 \begin{tabular}{l}
$\bullet$ $\rm{X}\rightarrow\rm{ZZ}\rightarrow\rm{(}\ell\ell\rm{)}\rm{(}\ell\ell\rm{)}$~\cite{Collaboration:2012iua}; \\
$\bullet$ $\rm{X}\rightarrow\rm{ZZ}\rightarrow\rm{(}\ell\ell\rm{)}\rm{(qq)}$~\cite{Collaboration:2012iua,Chatrchyan:2012pa,ChatrchyanABCDE}; \\
$\bullet$ $\rm{X}\rightarrow\rm{ZZ}\rightarrow\rm{(}\nu\nu\rm{)}\rm{(qq)}$~\cite{ChatrchyanABCDE}; \\
$\bullet$ $\rm{X}\rightarrow\rm{WW}\rightarrow\rm{(}\ell\nu\rm{)}\rm{(}\ell'\nu'\rm{)}$~\cite{Aad:2012gf} $\star$; \\
$\bullet$ $\rm{X}\rightarrow\rm{WW / ZZ }\rightarrow\rm{(qq)(qq)}\rightarrow\rm{dijet}$~\cite{CMS-PAS-EXO-11-095-DiVTagDijets} $\star$. \\
%\end{itemize}
\end{tabular}
\\ \\
Although the results of these searches have not yet been combined, the overall picture suggests that, using 7 TeV data, 
ATLAS and CMS experiments are able to exclude bulk gravitons with masses below approximately 800-900~GeV assuming \koverm$=1.0$. 
A selection of these analyses ($\star$) was presented at the conference and is also reported below.\\

\begin{figure}[htbp]
%\vspace*{7.0cm}
\begin{center}
%\special{psfile=pic2012_template_fig.ps voffset=-60 vscale=40
%hscale= 40 hoffset=10 angle=0}
    \begin{tabular}{cc}
    \psfig{figure=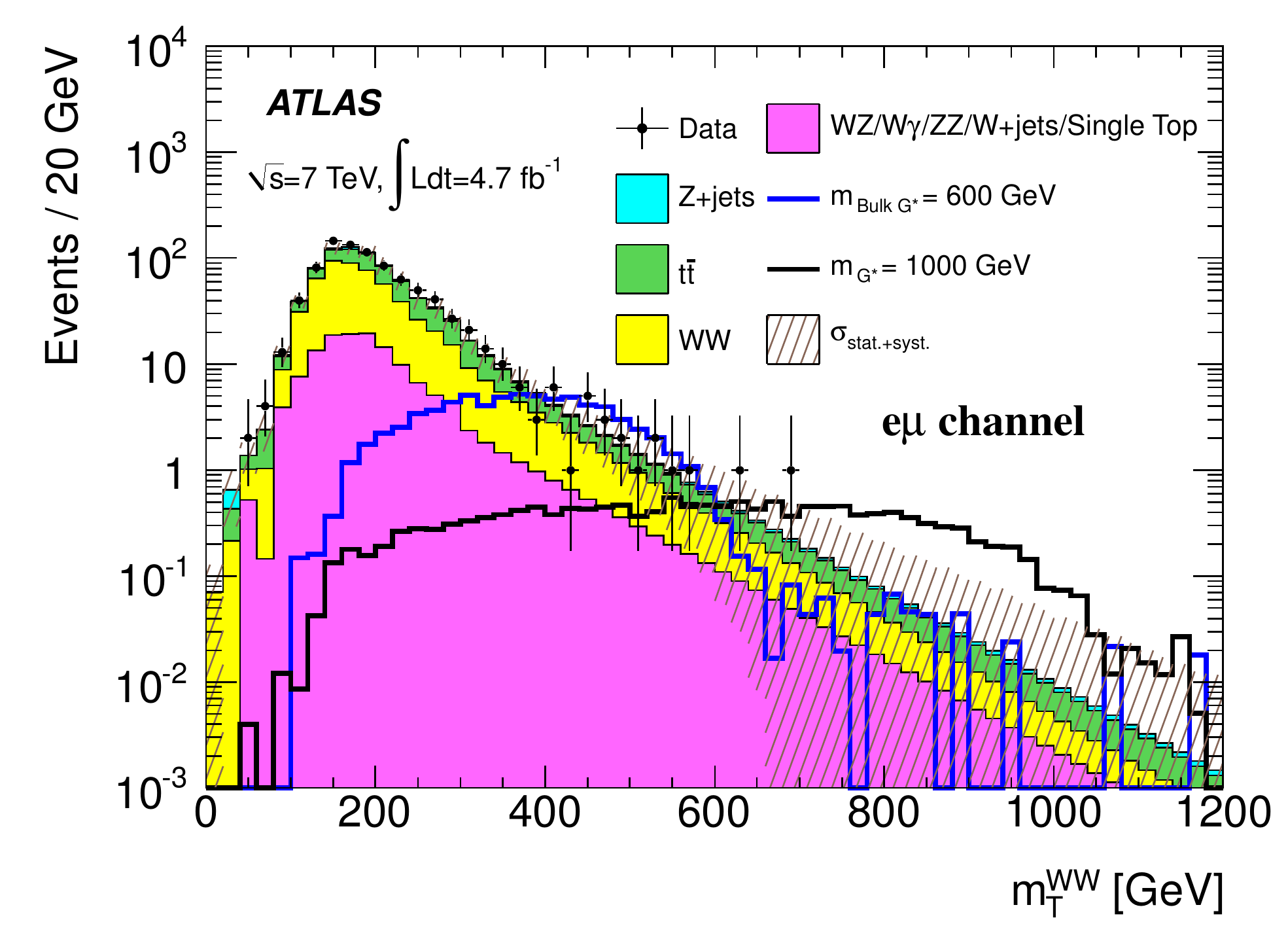,height=1.7in} &
    \psfig{figure=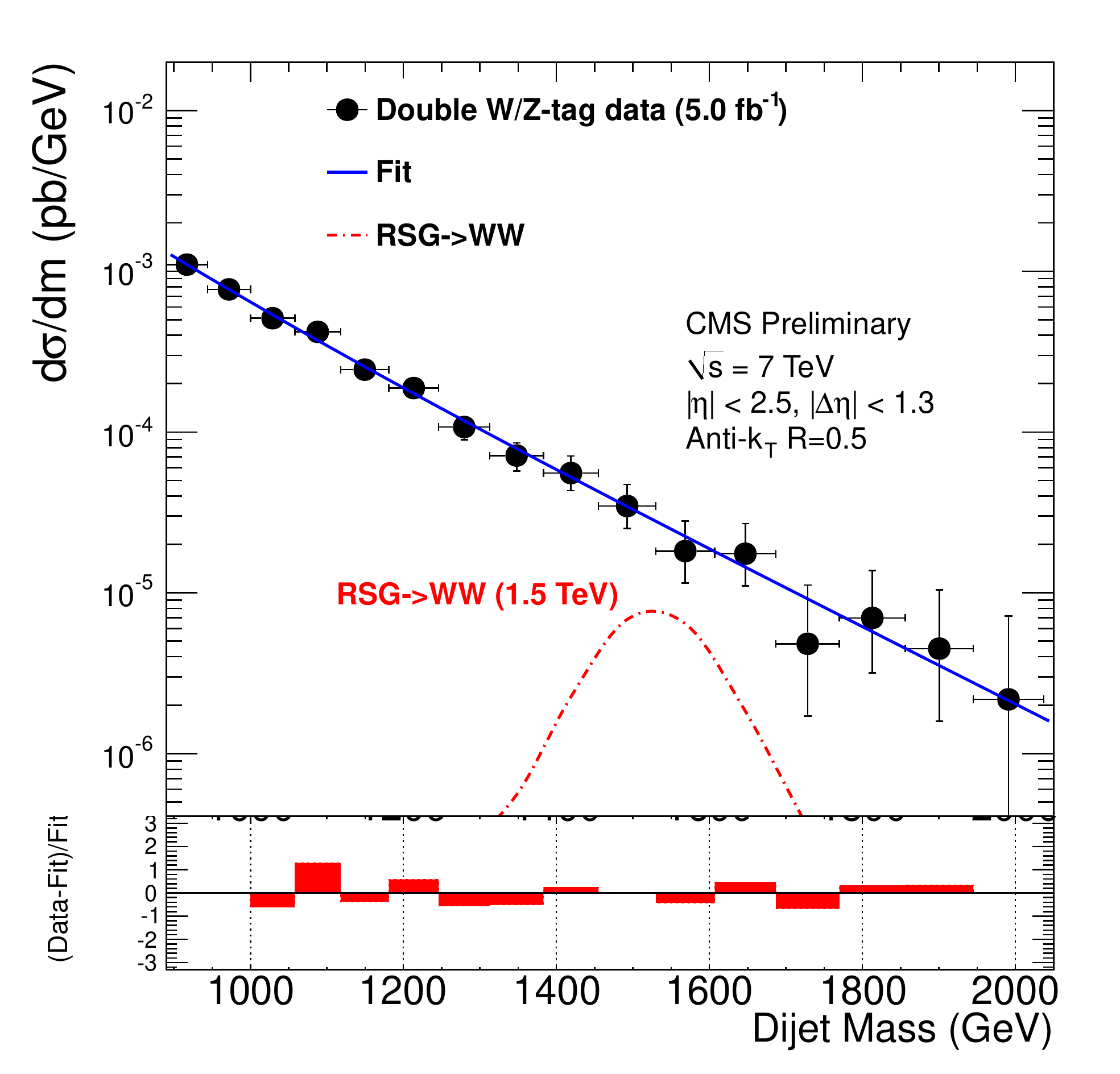,height=1.7in} \\
%\centerline{\epsfxsize=2.9in\epsfbox{BlakHoleST_CMS.pdf}} &
%\centerline{\epsfxsize=2.9in\epsfbox{BlakHoleST_CMS.pdf}} \\
    \end{tabular}
\caption[*]{(Left) Observed and predicted $\rm{m}_{\rm{T}}^{\rm{WW}}$ distribution after event selection in the e$\mu$ channel. 
Examples of RS1 and bulk graviton signals are also shown~\cite{Aad:2012gf}. (Right) Dijet mass spectrum for events in data with 
two identified W/Z-jets compared to background prediction. A graviton signal shape distribution with arbitrary cross section is also shown~\cite{CMS-PAS-EXO-11-095-DiVTagDijets}. }
\label{fig:lvlv_qqqq_RSGraviton}    
\end{center}
\end{figure}

A search for a heavy particle that decays to $\rm{WW}\rightarrow\rm{(}\ell\nu\rm{)}\rm{(}\ell'\nu'\rm{)}$ final states 
(ee, $\mu\mu$, e$\mu$) was performed by the ATLAS collaboration~\cite{Aad:2012gf}. No 
excess above the SM background prediction is observed 
in the transverse mass distribution of the WW system ($m_{\rm{T}}^{\rm{WW}}$) for events with 
two high \pt leptons and large \etmiss, as shown in Fig.~\ref{fig:lvlv_qqqq_RSGraviton} (left). 
Lower limits on the RS1 (bulk) graviton mass are set at 1.23 (0.84)~TeV, respectively, for \koverm$=0.1$ (1.0)
by combining the three dilepton channels.\\

A search for massive resonances decaying into a pair of vector bosons (WW, ZZ, or WZ) or into a quark and a vector boson 
(qW and qZ), where each vector boson decays in hadrons, has been performed by the CMS experiment~\cite{CMS-PAS-EXO-11-095-DiVTagDijets}. The analysis focus on resonances which are sufficiently heavy to result in boosted vector bosons; the decay products of each vector boson are then merged into a single jet (W/Z-jet), and the event effectively has a dijet topology. The analysis relies on recently developed techniques in the area of jet substructure and uses reconstructed 
quantities such as the pruned jet mass ($m_{jet}$) and the mass drop (defined as $\mbox{max[} m_{{\rm subjet1}}, m_{{\rm subjet2}}\mbox{]} / m_{jet}$) to identify W/Z-jets and suppress the large QCD dijet background. 
%The same techniques could also be used in the semi-leptonic channels $\rm{X}\rightarrow\rm{ZZ}\rightarrow\rm{(}\ell\ell\rm{)}\rm{(qq)}$ and $\rm{X}\rightarrow\rm{WW}\rightarrow\rm{(}\ell\nu\rm{)}\rm{(qq)}$.
 Figure~\ref{fig:lvlv_qqqq_RSGraviton} (right) 
 shows the dijet mass spectrum for events where each of the two jets is identified as a W/Z-jet; a dijet mass spectrum for events with only 
one identified W/Z-jet is also produced. 
No significant evidence for new resonance production in both dijet mass spectra is found. 
The sensitivity of this measurement with the present dataset is not sufficient to extract lower limits 
on the graviton mass for the models and couplings considered in the previous paragraph. 
Lower limits on the mass of excited quark resonances decaying into qW (qZ) at 2.32 (2.04) TeV are set instead. 
These are the most stringent limits in the qW and qZ final states to date.

\section{Long-Lived Particles}
A wide variety of theories beyond the SM allow for the possibility of new long-lived particles whose 
passage through the detector or whose decay in flight can be observed by the LHC experiments.
For instance, with Supersymmetry being strongly constrained by the experiments, 
there is growing interest in looking in great detail for SUSY long-lived particles that might 
have escaped the standard searches which assume prompt decays.
These exotic particles may be neutral, charged and/or colored. 

\subsection{Heavy Stable Charged Particles}
Heavy stable (or quasi-stable) charged particles (denoted as HSCP in this paper) 
appear in various extensions of the SM.  These particles have typically 
masses above 100~GeV thus implying low velocity ($\beta\gamma = p/m < \rm{ or } \ll 1$) given the typical momenta with which they are produced at LHC, where $\beta$ is the particle velocity in units of 
the speed of light ($c$), $\gamma = 1/\sqrt{1-\beta^2}$ is the Lorentz factor, $p$ is the particle momentum, and $m$ is the particle mass.
They are stable (or quasi-stable) with $\rm{c}\tau$ greater than the typical dimensions of an LHC detector ($>1-10$~m), where $\tau$ is the mean lifetime of the particle. 
They carry electric charge and therefore they can deposit energy 
in the detector via ionization ($dE/dx$ signature). 

Several searches for these exotic particles have been performed by the 
ATLAS and CMS experiments in 2011: searches for fractionally-charged HSCPs~\cite{CMS:2012xi}, unit-charge HSCPs~\cite{Aad:2012vd,Chatrchyan:2012sp} including HSCPs stopped in the detector volume~\cite{Chatrchyan:2012yg}, and multi-charged HSCPs~\cite{CMS-PAS-EXO-11-090-HSCPMultiCharge} including magnetic monopoles~\cite{Aad:2012qi}.
Overall the analyses rely on a few key elements to distinguish these particles from 
the quasi-stable SM particles produced in background processes: 
\begin{enumerate}
\item HSCPs are typically reconstructed as high \pt tracks in the inner tracker detector (and can reach the muon detectors in some cases);
\item low value of $\beta\gamma$ implies large $dE/dx$ in the inner tracker detectors for particles with charge $\ge 1$. HSCPs with fractional charge can be identified by an anomalous low value of $dE/dx$ in the inner tracker detectors due to the quadratic dependence of the charge in the ionization energy loss;
\item  low velocity $\beta$ also implies late arrival to the detectors from the interaction point, which can be identified by time of flight (TOF) measurements.
\end{enumerate}
A selection of searches for HSCPs was presented at the conference 
and the results are summarized below. \\

The CMS experiment has searched for heavy long-lived particles with hadronic nature, such as gluinos or stops, which hadronize in flight, forming meta-stable bound states with quarks and gluons (so called R-Hadrons)~\cite{Chatrchyan:2012sp}. The analysis is also sensitive to lepton-like HSCPs, such as staus in Gauge Mediated Supersymmetry Breaking (GMSB) models. 
The inner tracking detectors are used to define a sample of events containing tracks with high momentum and high 
ionization energy loss ($dE/dx$). A second sample of events with high-momentum and high-ionization tracks satisfying muon identification and long TOF criteria is analyzed independently. 
In both samples the results are consistent with the data-driven background estimates that exploit the absence of correlation among \pt, $dE/dx$, and TOF measurements (ABCD-like methods). Lower limits are set on the mass of long-lived gluinos, stops, and GMSB staus at 1098, 737, and 223 GeV, respectively. 
Limits are about 100 GeV weaker for gluinos and stops that hadronize into a neutral bound state before reaching the muon detectors.
%The mass of the HSCP candidates ($\rm{m}_{\rm{HSCP}}$) is measured by using an approximate Bethe-Bloch formula, $\rm{dE/dX} = \rm{K} \cdot \rm{(}\rm{m}_{\rm{HSCP}}^2/\rm{p}^2\rm{)}+C$, where K, C are determined from data using a control sample of low-momentum protons. 
Similar techniques are also used in the ATLAS analysis~\cite{Aad:2012vd} bringing to comparable 
sensitivity of the search. \\

The ATLAS experiment has performed the first dedicated search at LHC for magnetic monopoles~\cite{Aad:2012qi}.
%Magnetic monopoles have long been the subject of dedicated search efforts both in astrophysical sources and at colliders for three main reasons: their introuction into the theory of electromagnetism would restore the symmetry between electricity and magnetism in MaxwellÕs equations; their existence would explain the quantization of electric charge and they appear in many grand uniÞed theories. 
The Dirac quantization condition leads to a prediction for the minimum unit magnetic charge $g$: $g/e=1/2\alpha_{e}\approx 68.5$, where $e$ is the electric charge and $\alpha_{e}$ is the fine structure constant. In addition the trajectory of an electrically neutral magnetic monopole in the inner detector is straight in the $r-\phi$ plane and curved in the $r-z$ plane.

%Monopoles are highly ionizing particles interacting with matter like an ion of electric charge of 68.5$e$, thus producing a large number of $\delta-$rays in the transition radiation tracker of ATLAS along their trajectory.
Monopoles are identified in the ATLAS detector as high energy clusters in the electromagnetic (EM) calorimeter with an anomalous large ionization energy loss (comparable to that of an ion with electric charge of 68.5$e$) in the transition radiation tracker (TRT) along their trajectory. 
In addition magnetic monopoles give rise to a narrow ionization energy 
deposit in the EM calorimeter (since bremsstrahlung and $\mbox{e}^+\mbox{e}^-$ pair production are negligible), 
the size of which provides another powerful discriminator of the monopole signal from 
backgrounds such as electrons and photons, which induce an EM shower. No event is found in 
the signal region after the final event selection, which is compatible with the data-driven background expectation.

%The duality of Maxwell's equations implies a magnetic coupling that grows with the square of the velocity $\beta$. 
For relativistic monopoles the magnetic coupling is very large, thus precluding any perturbative 
calculation of monopole production processes at colliders.  Therefore, the main result of this analysis is a cross section upper limit of 2 fb for Dirac monopoles with the minimum unit magnetic charge with mass between 200 GeV and 1500 GeV for the fiducial region defined by i) pseudorapidity $|\eta|<1.37$ and ii) transverse kinetic energy 
$600-700 < E^{\rm{kin}}sin(\theta) < 1400$~GeV, derived without assuming a particular production mechanism. 
This is the first direct collider search that yields cross section constraints on magnetic monopoles with masses greater than 900 GeV.

\subsection{Long-Lived Neutral Particles}

ATLAS and CMS experiments have also performed several searches for long-lived neutral 
particles. These objects can be identified by reconstructing their displaced decay vertex (using
tracking and/or timing information). These experimental techniques require that the 
decay of the neutral long-lived particle happens within the detector volume, thus 
putting limits on the maximum lifetime that can be probed by these analyses 
(typically of the order of 1-10 ns). 
No sign of new physics is found in these searches, which are listed below for reference: 
\begin{itemize} 
\item searches for long-lived neutral particles decaying into a photon and invisible 
particles~\cite{Chatrchyan:2012ir,CMS-PAS-EXO-11-035-DisplacedPhotons};
\item searches for heavy resonances decaying to long-lived massive neutral particles that 
each decay to pairs of charged fermions~\cite{ATLAS:2012av,Aad:2012kw,CMS-PAS-EXO-11-101-DisplacedLeptons};
\item search for decay of heavy neutral particles producing a multi-track vertex with a muon and hadrons~\cite{Aad:2012zx}.
\end{itemize}

\section{Leptoquarks}

The Standard Model has an intriguing but ad hoc symmetry between 
quarks and leptons. In some theories beyond the SM, such 
as SU(5) gran unification, Pati-Salam SU(4), and others, the existence 
of a new symmetry relates the quarks and leptons in a fundamental way. 
These models predict the existence of new bosons, called leptoquarks. 
The leptoquark (LQ) is coloured, has fractional electric charge, and 
decays to a charged lepton and a quark with unknown branching 
fraction $\beta$, or a neutrino and a quark with branching fraction 
$(1-\beta)$. Constraints from experiments sensitive to flavour-changing 
neutral currents, lepton-family-number violation, and other rare processes 
favour LQs that couple to quarks and leptons within the same SM generation, 
for LQ masses accessible to current colliders. 

Searches for pair-production of first and second 
generation scalar LQs have been performed in the \eejj, 
\enujj, \mumujj, and \munujj channels~\cite{Aad:2011ch,ATLAS:2012aq,Chatrchyan:2012dn} 
by both ATLAS and CMS collaborations. 
%Figure~\ref{fig:Leptoquarks} (left) reports limits in the 
%plane $\beta$ vs second generation LQ mass. 
The most stringent lower limits to date on the mass of first and second 
generation LQs are 830 (640) and 840 (650), respectively, for $\beta=1$ (0.5).

\begin{figure}[htbp]
%\vspace*{7.0cm}
\begin{center}
%\special{psfile=pic2012_template_fig.ps voffset=-60 vscale=40
%hscale= 40 hoffset=10 angle=0}
    \begin{tabular}{cc}
    \psfig{figure=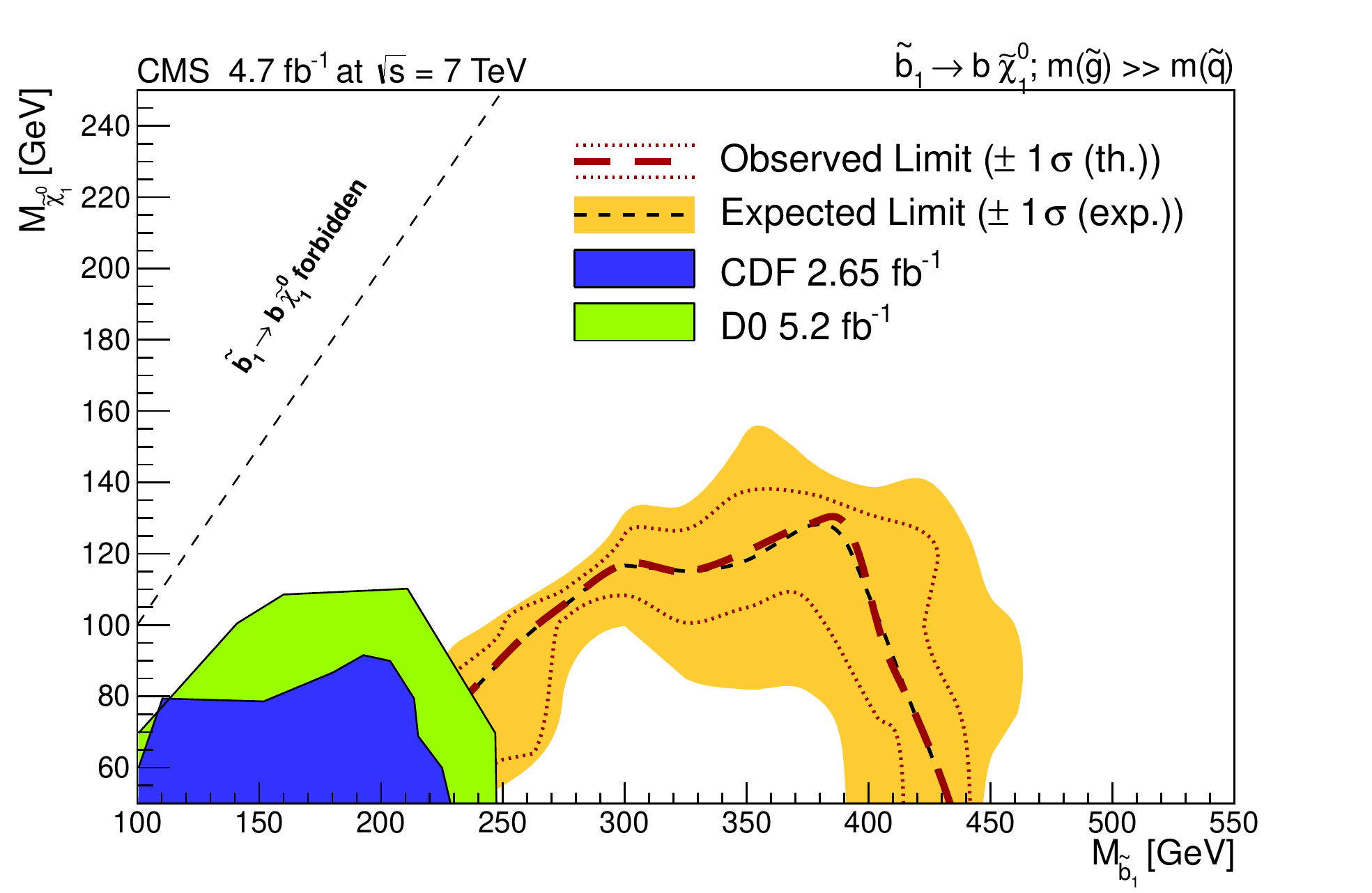,height=1.8in} &
    \psfig{figure=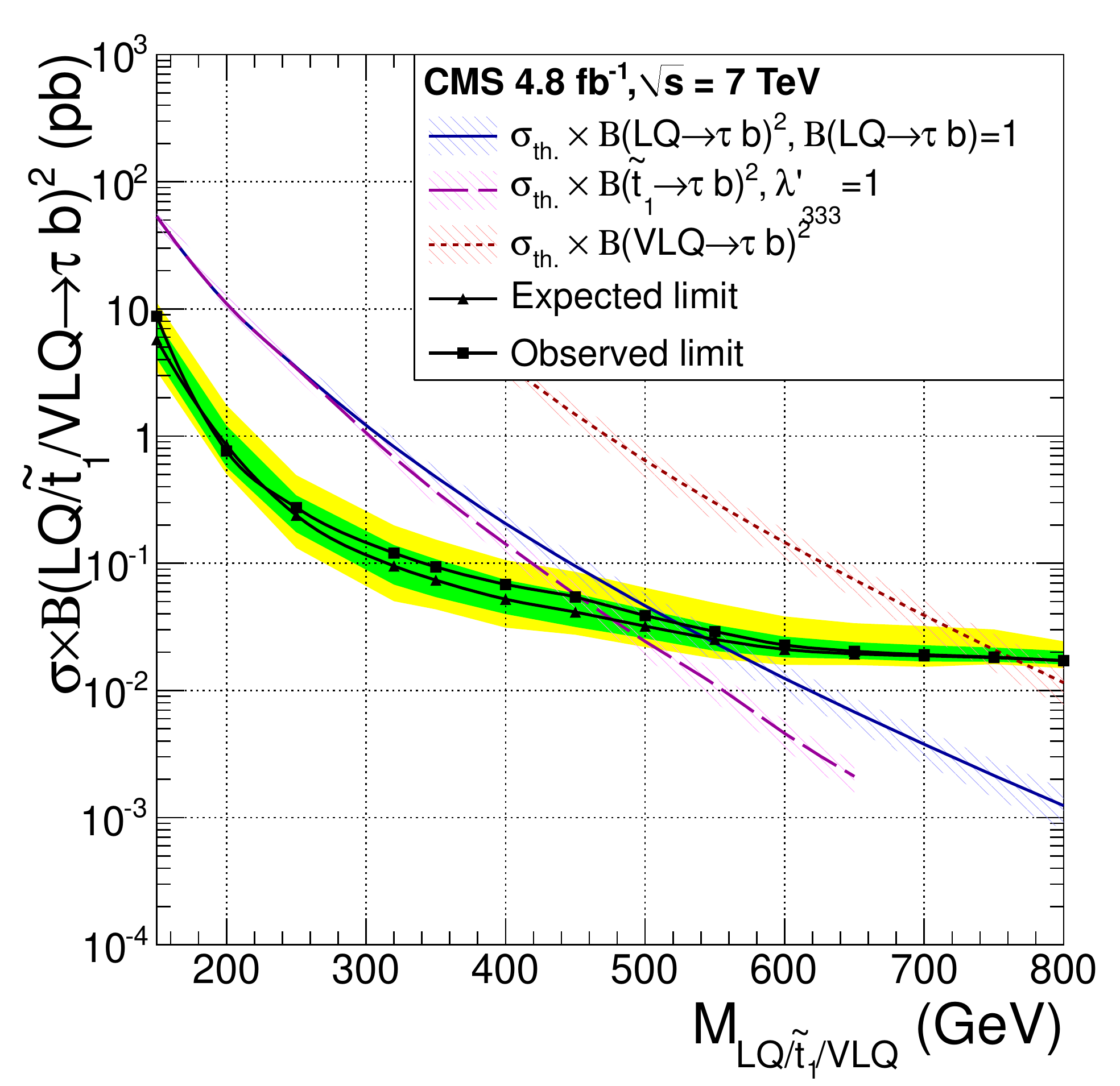,height=1.8in} \\
%\centerline{\epsfxsize=2.9in\epsfbox{BlakHoleST_CMS.pdf}} &
%\centerline{\epsfxsize=2.9in\epsfbox{BlakHoleST_CMS.pdf}} \\
    \end{tabular}
\caption[*]{Mass limits from \nunubb (left)~\cite{Chatrchyan:2012st}  and \tautaubb (right) CMS searches~\cite{Chatrchyan:2012sv}.}
\label{fig:Leptoquarks}    
\end{center}
\end{figure}

Searches for pair-production of third generation leptoquarks have been performed
by the CMS experiment in two final states: \nunubb~\cite{Chatrchyan:2012st} 
and \tautaubb~\cite{Chatrchyan:2012sv}.
The \nunubb analysis employs data-driven background estimates based 
on the {\sc Razor} variables, already used extensively in various jets+\etmiss SUSY searches at CMS. 
The \tautaubb analysis looks at final states with 2 reconstructed $\tau$ leptons and two b-tagged jets, 
where one $\tau$ decays in hadrons, while the other decays to either $\rm{e}\bar{\nu}_{\rm{e}}\nu_{\tau}$ or 
$\mu\bar{\nu}_{\mu}\nu_{\tau}$. Lower mass limits on third generation leptoquarks are set to 525 (450) GeV 
assuming $\beta=1$ (0). In addition both searches are also sensitive to pair production of 
bottom squarks and top squarks~\footnote{Assuming R-Parity violation decays of the stops.} 
predicted by SUSY scenarios, as shown in Figure~\ref{fig:Leptoquarks}.

%The CMS experiment has also searched for third generation LQs in the channels
%\tau\taujj~\cite{} and \bbnunu~\cite{}. 

\section{Conclusions} 
No sign of physics beyond the Standard Model has been found so far in many different final states 
using the 5 \invfb of data collected by ATLAS and CMS detectors in 2011 at $\sqrt{s}=7$~TeV. 
Searches for exotic new physics phenomena in 2012 at $\sqrt{s}=8$~TeV are currently ongoing;  
only a few preliminary results have been shown in conferences and 
found to be consistent with the Standard Model predictions. 
The larger integrated luminosity expected by the end of 2012 (about 20~\invfb) 
and the higher LHC energy compared to 2011 extend significantly the sensitivity 
at high mass for most of the searches presented in this review. 
Experiments are also looking into new final states not previously explored and the results 
of these analyses will be released in the next months. 
In conclusion, there is another interesting year ahead of us concerning searches 
for exotic, new physics phenomena before the long 2013-2014 LHC shutdown!

\section*{Acknowledgements} The author wishes to thank the organizers of "PIC2012" for the rich and interesting physics program, and the beautiful location of the conference.

%\section*{Appendix}
%This is place for Appendix, if any.

\end{document}